\documentclass[apj]{emulateapj}
\usepackage{mathptmx}

\def\gtorder{\mathrel{\raise.3ex\hbox{$>$}\mkern-14mu
             \lower0.6ex\hbox{$\sim$}}}
\def\ltorder{\mathrel{\raise.3ex\hbox{$<$}\mkern-14mu
             \lower0.6ex\hbox{$\sim$}}}

\slugcomment{accepted: July 21, 2006}

\shorttitle{The Short-Hard GRB~051103}

\shortauthors{Ofek et al.}

\begin{document}

\title{The Short-Hard GRB~051103: Observations and implications for its nature}
\author{
E.~O.~Ofek\altaffilmark{1},
S.~R.~Kulkarni\altaffilmark{1},
E.~Nakar\altaffilmark{1},
S.~B.~Cenko\altaffilmark{1},
P.~B.~Cameron\altaffilmark{1},
D.~A.~Frail\altaffilmark{2},
A.~Gal-Yam\altaffilmark{1},
A.~M.~Soderberg\altaffilmark{1},
and D.~B.~Fox\altaffilmark{3}
}
\altaffiltext{1}{Division of Physics, Mathematics and Astronomy, California Institute of Technology, Pasadena, California 91125, USA}
\altaffiltext{2}{National Radio Astronomy Observatory, PO Box 0, Socorro, New Mexico 87801, USA}
\altaffiltext{3}{Department of Astronomy and Astrophysics, 525 Davey Laboratory, Pennsylvania State University, University Park, Pennsylvania 16802, USA}

\begin{abstract}

The bright short-hard
GRB~051103 was triangulated by the inter-planetary network
and found to occur in the direction
of the nearby M81/M82 galaxy group.
Given its possible local-Universe nature,
we searched for an afterglow associated with this burst.
We observed the entire 3-$\sigma$ error quadrilateral using the Palomar 60-inch
robotic telescope and the Very Large Array (VLA)
about three days after the burst. 
We used the optical and radio observations to constrain the flux of
any afterglow related to this burst, and to show that this burst 
is not associated with a typical supernova 
out to $z\approx 0.15$.
Our optical and radio observations,
along with the Konus/Wind gamma-ray energy and light curve
are consistent with this burst being a giant flare
of a Soft Gamma-ray Repeater (SGR)
within the M81 galaxy group.
Furthermore, we find a star forming region
associated with M81 within the error quadrilateral of this
burst which supports the SGR hypothesis.
If confirmed, this will be the first case of a
soft gamma-ray repeater outside the local group.

\end{abstract}

\keywords{
gamma rays: bursts: individual: GRB~051103 ---
stars: neutron ---
galaxies: individual (M81, M82)}

\section{Introduction}
\label{Introduction}

The recent discovery of short-hard Gamma-Ray Bursts (GRB)
afterglows in the X-ray (Gehrels et al. 2005; Fox et al. 2005),
optical (Hjorth et al. 2005), and radio (Berger et al. 2005)
demonstrates that short-hard GRBs reside in both early- and late-type
galaxies
with at least some preference for old systems
(e.g. Gal-Yam et al. 2005; Tanvir et al. 2005)
and release energy of the order of $10^{50}$~erg.
With the detection of the giant flare of
the 2004 December 27, from 
the Soft Gamma-ray Repeater
(SGR)~1806$-$20
(e.g. Hurley et al. 2005),
it has been suggested
that a fraction (or all) of the short-hard bursts population
originates from giant flares of SGR
in nearby galaxies (e.g. Dar 2005; Hurley et al. 2005).
This idea was based on the observed Galactic rate
of SGR giant flares and the star formation rate
in the local Universe.
However, the fraction of extragalactic SGR flares among
short-hard GRBs was shown to be small
(Nakar et al. 2006; Gal-Yam et al. 2005;
Palmer et al. 2005; Popov \& Stern 2005;
Lazzati et al. 2005; Ofek 2006).
In contrast with the evidence associating
short-hard GRBs with old stellar population (e.g. Nakar et al. 2005a),
SGRs seem to
emerge from young population objects
(e.g. Gaensler et al. 2001; see however Levan et al. 2006;
for a recent review on SGRs see Woods \& Thompson 2006).

\begin{figure}
\centerline{\includegraphics[width=8.5cm]{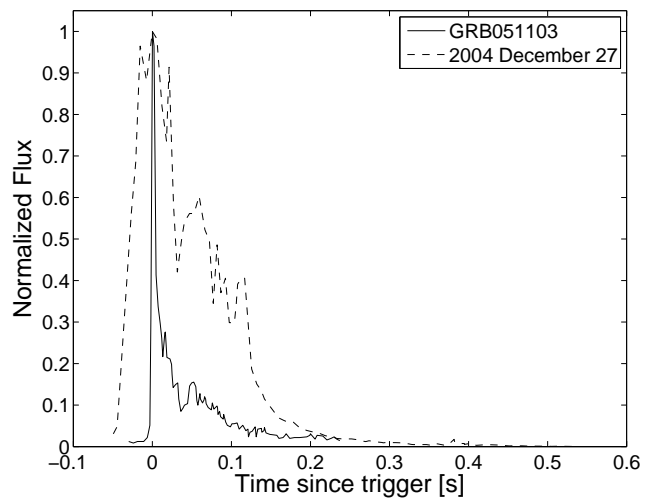}}
\caption{The Konus/Wind gamma-ray light curve of GRB~051103 (solid line),
compared with the light curve of the 2004 December 27 SGR giant flare (dashed line).
The light curve of the 2004 December 27 SGR giant flare is based on a digitization
of Figure~1 in Terasawa et al. (2005), while the light curve
of GRB~051103 is based on a digitization of the 18-1160 keV-band
light curve from the Konus/Wind
website.
%\footnote{http://www.ioffe.rssi.ru/LEA/GRBs/GRB051103\_T33943/}.
\label{GRB051103_LC_comparison}}
\end{figure}

On UTC 2005 November 3 09:25:43.785, a short-hard GRB
with $0.17\,$s duration was detected (Golenetskii et al. 2005)
by five satellites, of the Inter-Planetary Network
(IPN; e.g. Hurley et al. 1999),
carrying gamma-ray detectors:
Konus/Wind; HETE/Fregate; Mars Odyssey/GRS; Mars Odyssey/HEND;
RHESSI; and Swift/BAT.
The GRB fluence in the Konus/Wind $20$~keV-$10$~MeV band was
$2.34_{-0.28}^{+0.31}\times10^{-5}$~erg~cm$^{-2}$,
and its peak flux on two-milliseconds time scale was
$1.89_{-0.35}^{+0.25}\times10^{-3}$~erg~cm$^{-2}$~s$^{-1}$
($90\%$~confidence).
This peak flux was one of the largest ever observed for
Konus-Wind short GRBs, second only to GRB~031214 (Hurley et al. 2003).
Moreover,
comparing the burst fluence with the distribution of the integrated four-channel fluence
of all BATSE short-hard bursts, we find it to be in the $99$th
percentile.
The light curve, shown in Figure~\ref{GRB051103_LC_comparison} (solid line),
has a very steep rise on $\sim4$~milliseconds
time scale and a weak decaying tail
on time scale of $0.1\,$s.
The IPN 3-$\sigma$ error box, totaling 260 square arcminutes,
%(and not 120 square arcmin as stated in Golenetskii 2005)
includes the outskirts of the nearby galaxies M81 and M82.
At first glance, the error region seems
to exclude the main body of M81 or M82
(see however \S\ref{Obs}).

In this paper, we present new optical and radio observations of M81 and M82
region
which cover the entire IPN error quadrilateral, taken several days
after the burst, and use them to put limits on any
afterglow emission related to this GRB.
We combine these with Galaxy Evolution Explorer (GALEX)
archival data,
and discuss the implications for the nature of this burst.

\section{Observations}
\label{Obs}

\begin{figure}
\centerline{\includegraphics[width=8.5cm]{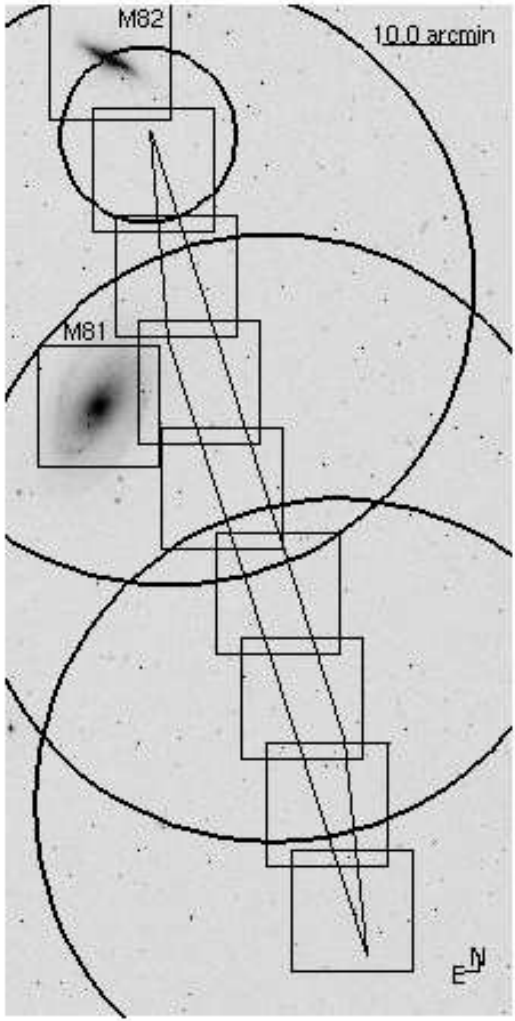}}
\caption {The IPN error quadrilateral of GRB~051103 plotted
over the Palomar Sky Survey II R-band image.
The boxes mark the Palomar 60-inch field of view for the 
ten pointings (Table~\ref{Tab-Log}):
eight pointings which cover the IPN error quadrilateral and two
pointings around M81 and M82.
The circles mark the four VLA pointings (see \S\ref{Obs}).
The coordinates of the points defining the
IPN error quadrilateral, from south to north, are:
09$^{h}$50$^{m}$16.$^{s}$8 $+$68$^{\circ}$06$^{'}$54$^{''}$;
09$^{h}$50$^{m}$55.$^{s}$4 $+$68$^{\circ}$29$^{'}$56$^{''}$;
09$^{h}$54$^{m}$15.$^{s}$1 $+$69$^{\circ}$11$^{'}$20$^{''}$;
09$^{h}$54$^{m}$57.$^{s}$1 $+$69$^{\circ}$33$^{'}$50$^{''}$ (J2000.0).
\label{ErrorBox_DSS_R} }
\end{figure}

Figure~\ref{ErrorBox_DSS_R} shows the
IPN error quadrilateral of GRB~051103, plotted over the 
Palomar Sky Survey (POSS) II R-band image of this field.
The error quadrilateral is the elongated box
running from  south-west to north-east.
We observed the IPN error quadrilateral
using the Palomar 60-inch robotic telescope, equipped with
a $2$k$\times2$k SITe CCD with a pixel scale of $0\farcs378\,$pix$^{-1}$,
in R-band under good conditions.
We acquired eight pointings which cover
the error quadrilateral and two pointings centered on M81 and M82.
The field of view of the pointings are marked as
boxes in Fig.~\ref{ErrorBox_DSS_R}.
\begin{deluxetable}{lccc}
\tablecolumns{4}
%\tabletypesize{\footnotesize}
%\tabletypesize{\scriptsize}
\tablewidth{0pt}
\tablecaption{Log of P60 observations}
\tablehead{
\colhead{Field Name} &
\colhead{R.A. J2000.0 Dec.} &
\colhead{Date} &
\colhead{Exp. Time} \\
\colhead{} &
\colhead{} &
\colhead{} &
\colhead{[s]}
}
\startdata
1   & 09:50:36.0 +68:11:42 & 2005-11-06.390  & 900  \\
    &                      & 2005-11-15.351  & 540  \\
2   & 09:51:12.0 +68:22:48 & 2005-11-06.397  & 900  \\
    &                      & 2005-11-15.384  & 540  \\
3   & 09:51:48.0 +68:33:54 & 2005-11-06.407  & 900  \\
    &                      & 2005-11-15.389  & 540  \\
4   & 09:52:24.0 +68:45:00 & 2005-11-06.419  & 900  \\
    &                      & 2005-11-15.406  & 540  \\
5   & 09:53:00.0 +68:56:06 & 2005-11-06.462  & 1800 \\
    &                      & 2005-11-12.343  & 540  \\
6   & 09:53:36.0 +69:07:12 & 2005-11-06.484  & 1620 \\
    &                      & 2005-11-12.346  & 540  \\
7   & 09:54:12.0 +69:18:18 & 2005-11-06.486  & 540  \\
    &                      & 2005-11-12.360  & 540  \\
8   & 09:54:48.0 +69:29:24 & 2005-11-06.498  & 540  \\
    &                      & 2005-11-12.367  & 540  \\
M81 & 09:55:33.2 +69:03:55 & 2005-11-08.374  & 900  \\
    &                      & 2005-11-15354   & 900  \\
M82 & 09:55:52.2 +69:40:49 & 2005-11-07.352  & 1080 \\
    &                      & 2005-11-15.339  & 900  \\
\enddata
\tablecomments{The typical seeing during these nights was about $2''$,
except on November 15th in which it was about $3''$.
The Moon phase on November 6th was $24\%$ and on November 15th it was $100\%$.}
\label{Tab-Log}
\end{deluxetable}
The log of observations is presented in Table~\ref{Tab-Log}.
The first epoch images of the error quadrilateral
(pointings one to eight) were obtained
$2.96$ to $3.16$~days after the GRB trigger.
We note that GRB~051103 was announced about 2.4~d
after the GRB occured.
Comparison of the two-epochs
by image-blinking and image-subtraction using ISIS (Alard \& Lupton 1998),
did not reveal any optical transient to a limiting magnitude of $R=20.5$
within the error quadrilateral and $R=19.0$ within
the cores of M81 and M82.
We clearly detected, however, one variable point source
within the error quadrilateral.
The source at $09^{h}53^{m}18.^{s}92$~$+69^{\circ}03'47\farcs5$
is cataloged by
Perelmuter \& Racine (1995; $V=15.15$; $B-V=0.75$; $V-R=0.57$).
As this detection is due to variability of a previously cataloged source, we
do not consider it interesting for our purposes.

We observed the field of GRB~051103 with the Very Large Array\footnote{The Very Large Array is operated by the National Radio
Astronomy Observatory, a facility of the National Science Foundation
operated under cooperative agreement by Associated Universities, Inc.}
(VLA)
in its most compact (D) configuration. All observations were
taken in standard continuum observing mode with a band width of
$2\times50$~MHz. We used 3C~48 for flux calibration, and for
phase referencing we used the calibrator J0949$+$662. Data were
reduced using standard packages within the Astronomical Image
Processing System (AIPS).
The first epoch observation,
beginning UT 2005 November $6.3$,
was conducted at a frequency of $1.4$~GHz.
We imaged the entire IPN error region
with three VLA pointings,
which are marked as
large circles in Fig.~\ref{ErrorBox_DSS_R}.
We found no new sources to the limit of the
NRAO VLA Sky Survey (NVSS; Condon et al. 1998).
The rms noise varied from $0.5$~mJy at the southern edge of the IPN error
quadrilateral, to $1.5$~mJy at the northern end
(due to residual flux from M82).
We also observed a $5$-arcminutes radius around
the southern part of M82, marked as a small circle
in Fig.~\ref{ErrorBox_DSS_R}, on UT 2005 November 8.45
at a frequency of 4.86 GHz. A comparison between this image
and the NVSS image of the same region did not reveal any new source
above 1.5 mJy (3-$\sigma$).

\begin{figure*}
\centerline{\includegraphics[width=17cm]{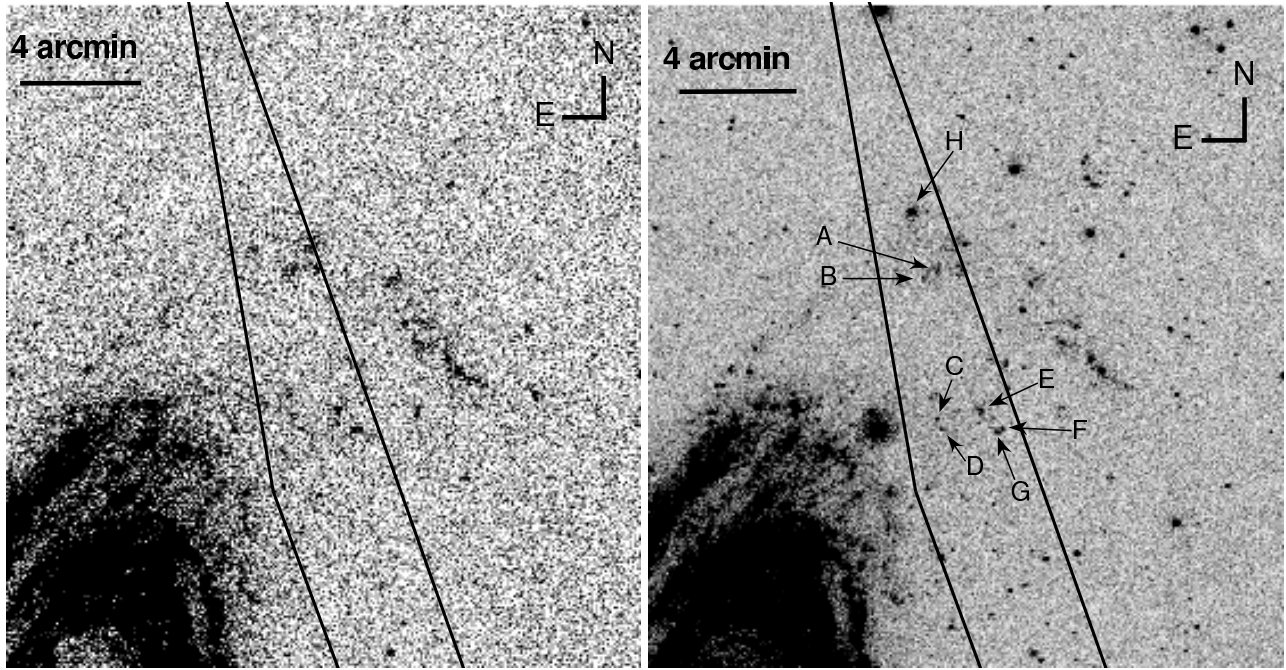}}
\caption {The GALEX FUV (left) and NUV (right) images of a region near
M81 containing UV-emitting objects.
The IPN error quadrilateral of GRB~051103 is overlayed.
Several of the brightest UV sources within the error quadrilateral
are marked and their UV-optical fluxes are listed in Table~\ref{Tab-UV}.
\label{GALEX_M81} }
\end{figure*}
\begin{deluxetable}{lcccccc}
\tablecolumns{7}
%\tabletypesize{\footnotesize}
%\tabletypesize{\scriptsize}
\tablewidth{0pt}
\tablecaption{Selected UV sources within the error quadrilateral}
\tablehead{
\colhead{Name} &
\colhead{R.A. J2000.0 Dec.} &
\colhead{FUV\tablenotemark{a}} &
\colhead{NUV\tablenotemark{a}} &
\colhead{B\tablenotemark{b}} &
\colhead{R\tablenotemark{b}} \\
\colhead{} &
\colhead{} &
\colhead{$\mu$Jy} &
\colhead{$\mu$Jy} &
\colhead{$\mu$Jy} &
\colhead{$\mu$Jy}
}
\startdata
A   &  09:54:06.52  $+$69:18:34.0 & $37.6\pm1.9$ &  $42.9\pm1.0$ &    70 &    70\\
B   &  09:54:11.05  $+$69:18:18.9 & $16.6\pm1.2$ &  $23.2\pm0.8$ &   340 &    10\\
C   &  09:54:06.66  $+$69:13:43.0 & $5.6\pm0.6$ &  $15.0\pm0.5$ &    20 &    30\\
D   &  09:54:04.76  $+$69:13:23.7 & $10.3\pm1.0$ &  $13.2\pm0.7$ &       &      \\
E   &  09:53:50.69  $+$69:14:00.0 & $35.4\pm1.7$ &  $35.6\pm0.9$ &       &      \\
F\tablenotemark{c}   &  09:53:43.25  $+$69:13:23.1 & $14.0\pm1.0$ &  $12.9\pm0.5$ &    20 &    30\\
G   &  09:53:45.30  $+$69:13:17.5 & $14.1\pm1.0$ &  $17.5\pm0.6$ &    20 &      \\
H\tablenotemark{d} &  09:54:15.33  $+$69:20:26.9 & $4.3\pm0.6$ & $451.4\pm3.3$ & 23300 & 24000\\
\enddata
\tablenotetext{a}{GALEX FUV/NUV specific flux.
To correct for Galactic extinction the NUV flux should be multiplied by 1.75
(Cardelli, Clayton, \& Mathis 1989; Schlegel, Finkbeiner, \& Davis 1998).}
\tablenotetext{b}{Approximate optical B/R-band specific flux from the USNO-B1.0 catalog (Monet et al. 2003).}
\tablenotetext{c}{Globular cluster in M81 (Object number 70319 in Perelmuter \& Racine 1995).}
\tablenotetext{d}{High proper motion star: GPM 148.564612+69.340917.}
\tablecomments{Selected UV sources within the error quadrilateral of GRB~051103,
found near the extension of the northern spiral arm of M81.
}
\label{Tab-UV}
\end{deluxetable}
M81 was the subject of several detailed studies.
For example: Matonick \& Fesen (1997) conducted an optical search for
supernova remnants;
Petit et al. (1988) and Lin et al. (2003) searched 
for HII regions in M81 and measured their properties;
and recently P\'{e}rez-Gonz\'{a}lez (2006) conducted a multi-wavelength study
of star formation in M81. However, none of these studies cover
the IPN error quadrilateral.
To search for possible star forming regions in M81/M82
that coincides with the IPN error quadrilateral,
we inspected the GALEX UV images of M81 and M82
(Hoopes et al. 2005).
Figure~\ref{GALEX_M81} shows the GALEX
near UV (NUV) and far UV (FUV) images, with the 
IPN error quadrilateral overlayed.
A region containing several UV-bright knots,
in an extension of the northern arm of M81,
is clearly seen within the error quadrilateral.
Several of the brightest UV sources,
found within the error quadrilateral,
are marked by arrows and their UV and optical fluxes are listed
in Table~\ref{Tab-UV}.
Some of these sources are extended
(in the POSS images) and blue, and may be
a young star forming regions in the outskirts of M81.
Assuming they are at the distance of M81,
the brightness of each of these sources in
the Near UV is equivalent to (at least) several tens 
of young O stars.

\section{Discussion}
\label{Disc}

The IPN error quadrilateral of the bright
GRB~051103 includes the outskirts of
the nearby ($3.63\pm0.34$~Mpc; Freedman et al. 1994)
galaxies M81 and M82 which are among the ten optically-brightest
galaxies in the sky.

Assuming that GRB~051103 is related to the M81/M82 group, and using a
distance modulus of $27.8$ (Freedman et al. 1994), and $A_R=0.2$ Galactic
extinction (Schlegel et al. 1998),
our null detection of any optical transients implies an upper limit of $-7.5$
on the R-band absolute magnitude of an optical transient.
This limit is comparable to the absolute magnitude of novae,
that peaks at $M_{V}=-6$ to $-10$ (e.g. della Valle \& Livio 1995)
and rule out the possibility that this event is associated with
a typical supernova (unless $z\gtorder0.15$, which
rules out a source in the local Universe).
The optical limit also rules out some
of the macronovae models
(assuming the distance of M81)
recently explored by Kulkarni (2005).

Given the observational properties of GRB~051103, two questions arises:
is it associated with the M81/M82 group?
And is it an SGR flare or ``genuine'' short-hard GRB?
Below, we explore the different possibilities and confront them
with the observational facts.

\subsection{An SGR in M81}

Assuming GRB~051103 was originated in M81,
the energy from this burst,
$(3.7\pm0.8)\times10^{46}$~erg
(including the uncertainty in distance to M81; Freedman et al. 1994),
is comparable to the energy of the most luminous SGR giant flare
yet detected.
For comparison, the isotropic energy release of the 1979 March 5
SGR flare was
$>6\times10^{44}$~erg, that
of 1998 August 27 flare was $2\times10^{44}$~erg,
while the energy release from the 2004 December 27
giant flare was as much as
$(3.7\pm0.9)\times10^{46}d_{15}^{2}$~erg (Hurley et al. 2005; Palmer et al. 2005),
where $d_{15}$ is the distance to SGR~1806$-$20 in $15$~kpc units.
As shown in Fig.~\ref{GRB051103_LC_comparison},
the gamma-ray light curve of this burst has a rise-time of about a millisecond,
which is characteristic of SGR giant flares (as well as less energetic
flares). Moreover, it is composed of a single major peak -- similar to other
SGR flares. Note that many, but not all, short-hard GRB light curves
show a more complex temporal
structure as observed in BATSE bursts (Nakar \& Piran 2002),
as well as in recent Swift bursts (e.g. GRB~051221; Parsons et al. 2005).

Known SGRs are associated with star-forming regions
(Gaensler et al. 2001).
Therefore, if GRB~051103 was an SGR giant flare
in M81 or M82, then we expect it to occur within a young
star forming region.
Indeed, we have found UV sources indicating a young stellar
population, in an extension of the M81 northern spiral arm,
within the IPN error quadrilateral
(see Fig.~\ref{GALEX_M81} and Table~\ref{Tab-UV}).

The SGR-origin idea is consistent with our
radio observational limits.
An extrapolation of the $1.4$~GHz specific flux of
SGR~1806$-$20 (Cameron et al. 2005)
to three days after the burst, suggests that
the specific flux of a 2004 December 27-like event in M81,
three days after the GRB,
would be $\sim0.02$~d$_{15}^{-2}$~mJy, significantly below
our VLA limit.

The gamma-ray spectrum is an important clue for the nature of such bursts
(e.g. Lazzati, Ghirlanda, \& Ghisellini 2005).
Unfortunately, the gamma-ray spectrum of GRB~051103
was not published.
Therefore, it is not clear if the spectrum
can be described by a black body spectrum,
as observed in the 2004 December 27 giant flare.
Assuming the spectrum is consistent with that
of a black-body,
the peak energy of GRB~051103,
$1.92\pm0.40$~MeV (Golenetskii et al. 2005),
is translated to a black body peak energy ($3kT$)
of $0.64\pm0.13$~MeV.
Given the peak flux (on a two-milliseconds time scale),
and the distance to M81,
the peak luminosity of GRB~051103
is about $2\times10^{48}$~erg~s$^{-1}$.
%Interestingly,
The black-body radius inferred
from this luminosity is $12\pm4$~km.
This radius is comparable to the radius
inferred for the 2004 December 27 giant flare,
obtained in a similar way
(Hurley et al. 2005; Nakar, Piran, \& Sari 2005b).
Intriguingly, these radii are comparable to
the typical radius of neutron stars.

\subsection{Short-hard GRB in the background}

Another possibility, already suggested by Lipunov et al. (2005), 
is that GRB~051103 is
actually a genuine short-hard GRB.
In this case, given the observed energy release from other short-hard GRBs
(e.g. $\sim10^{50}$~erg; Berger et al. 2005), this burst is more likely
a background event unrelated to M81/M82.
If however, it was a genuine short-hard GRB in M81
then we need to invoke a fainter population of short-hard GRBs,
for which the implied rate will be higher than
current estimates (e.g. Nakar et al. 2006).

Scaling the radio and optical properties of the afterglow of the
relatively well observed
short-hard GRB~050724 (Berger et al. 2005),
by the ratio of gamma-ray fluence of GRB~051103 to GRB~050724,
we estimate that
at the time of our VLA observation,
GRB~051103 should have had a $\sim40$~mJy
radio afterglow (at 8.46~GHz) -- an order of magnitude above our
detection limit.
Moreover, the non-detection of an optical afterglow,
assuming a power-law decay,
implies that the optical decay power-law was $\ltorder-2$.
However, the afterglow properties of short-hard GRBs
are non-homogeneous
(e.g. Berger et al. 2005; Hjorth et al. 2005; Bloom et al. 2006).
Therefore, our observations can not rule out the possibility
that this event was a genuine short-hard GRB.

If GRB~051103 was a genuine short-hard GRB in the background of M81,
then we expect the spectrum to be non-thermal.
In this case,
the fact that this GRB is optically thin to gamma-ray photons
allows to put a lower limit on its Lorentz factor
(i.e. the ``compactness problem'').
We used the recipe of Lithwick \& Sari (2001)
to estimate a lower limit on the Lorentz factor, $\gamma$.
We assumed that the gamma-ray spectrum of GRB~051103
is described by a Band spectrum (Band et al. 1993),
and used the duration of the event ($\delta{T}=0.17$~s),
the peak energy, and the fluence within the Konus/Wind spectral-band.
Assuming the gamma-ray spectrum of GRB~051103
can indeed be described by a Band-spectrum,
and the distance to the GRB is $3.6$~Mpc (500~Mpc)
we can set a lower limit of $\gamma\gtorder12$ ($\gtorder18$), on
the Lorentz factor of GRB~051103.

To conclude,
although we can not rule out the possibility that GRB~051103
was a genuine short-hard GRB,
the SGR giant flare appears to be the simplest interpretation
of this event.
The decay time, the peak luminosity and isotropic energy release,
the presence of UV sources in the error quadrilateral, and the lack of optical
and radio afterglow are all consistent with this burst being
a giant flare from an SGR in M81.
The spectrum of GRB~051103 has not been published.
If a future analysis of the gamma-ray spectrum
of GRB~051103 will show it is consistent with a black body
spectrum, this will be another important
evidence in favor of the SGR origin of this burst.
%
%We note that, although the rate of SGR hyper flares
%(i.e. with energy above $4\times10^{46}$~erg)
%is low ($[0.7-5]\times10^{-4}$ yr$^{-1}$ per SGR; Ofek 2006),
%SGR activity is highly correlated
%(Laros et al. 1987; Hurley et al. 1994; Cheng et al. 1996),
%and therefore the probability of detecting another
%hyper flare from the same SGR
%is considerably higher than naive estimates (Ofek 2006).
%

\acknowledgments
This work is supported in part by grants from NSF and NASA.
We are grateful to an anonymous referee for his useful comments.


\begin{thebibliography}{}



\bibitem[Alard \& Lupton(1998)]{1998ApJ...503..325A} Alard, C., \& Lupton, 
R.~H.\ 1998, ApJ, 503, 325 

\bibitem[Band et al.(1993)]{1993ApJ...413..281B} Band, D., et al.\ 1993, 
ApJ, 413, 281 

%\bibitem[Barthelmy et al.(2005)]{2005Natur.438..994B} Barthelmy, S.~D., et 
%al.\ 2005, Nature, 438, 994 
 
\bibitem[Berger et al.(2005)]{2005Natur.438..988B} Berger, E., et al.\ 
2005, Nature, 438, 988 

\bibitem[Bloom et al.(2006)]{2006ApJ...638..354B} Bloom, J.~S., et al.\ 
2006, ApJ, 638, 354 
 
\bibitem[Cameron et al.(2005)]{2005Natur.434.1112C} Cameron, P.~B., et al.\ 
2005, Nature, 434, 1112

\bibitem[Cardelli et al.(1989)]{1989ApJ...345..245C} Cardelli, J.~A., 
Clayton, G.~C., \& Mathis, J.~S.\ 1989, ApJ, 345, 245 

\bibitem[Cheng et al.(1996)]{1996Natur.382..518C} Cheng, B., Epstein, 
R.~I., Guyer, R.~A., \& Young, C.\ 1996, Nature, 382, 518 

\bibitem[Condon et al.(1998)]{1998AJ....115.1693C} Condon, J.~J., Cotton, 
W.~D., Greisen, E.~W., Yin, Q.~F., Perley, R.~A., Taylor, G.~B., \& 
Broderick, J.~J.\ 1998, AJ, 115, 1693 

\bibitem[Dar(2005)]{2005GCN..2942....1D} Dar, A.\ 2005, GRB Coordinates 
Network, 2942, 1 

\bibitem[della Valle \& Livio(1995)]{1995ApJ...452..704D} della Valle, M., 
\& Livio, M.\ 1995, ApJ, 452, 704 

\bibitem[Fox et al.(2005)]{2005Natur.437..845F} Fox, D.~B., et al.\ 2005, 
Nature, 437, 845 

\bibitem[Freedman et al.(1994)]{1994ApJ...427..628F} Freedman, W.~L., et 
al.\ 1994, ApJ, 427, 628 

\bibitem[Gaensler et al.(2001)]{2001ApJ...559..963G} Gaensler, B.~M., 
Slane, P.~O., Gotthelf, E.~V., \& Vasisht, G.\ 2001, ApJ, 559, 963 

\bibitem[Gal-Yam et al.(2005)]{2005astro.ph..9891G} Gal-Yam, A., et al.\ 
2005, astro-ph/0509891 

\bibitem[Gehrels et al.(2005)]{2005Natur.437..851G} Gehrels, N., et al.\ 
2005, Nature, 437, 851 

\bibitem[Golenetskii et al.(2005)]{2005GCN..4197....1G} Golenetskii, S., et 
al.\ 2005, GRB Coordinates Network, 4197, 1 

\bibitem[Hjorth et al.(2005)]{2005Natur.437..859H} Hjorth, J., et al.\ 
2005, Nature, 437, 859

\bibitem[Hoopes et al.(2005)]{2005ApJ...619L..99H} Hoopes, C.~G., et al.\ 
2005, ApJL, 619, L99 

\bibitem[Hurley et al.(1994)]{1994A&A...288L..49H} Hurley, K.~J., McBreen, 
B., Rabbette, M., \& Steel, S.\ 1994, A\&A, 288, L49  

\bibitem[Hurley et al.(1999)]{1999ApJS..120..399H} Hurley, K., Briggs, 
M.~S., Kippen, R.~M., Kouveliotou, C., Meegan, C., Fishman, G., Cline, T., 
\& Boer, M.\ 1999, ApJS, 120, 399 

\bibitem[Hurley et al.(2003)]{2003GCN..2492....1O} Hurley, K., et al. 2003, GRB Coordinates Network, 2492, 1 

\bibitem[Hurley et al.(2005)]{2005Natur.434.1098H} Hurley, K., et al.\ 
2005, Nature, 434, 1098 

\bibitem[Kulkarni(2005)]{2005astro.ph.10256K} Kulkarni, S.~R.\ 2005, astro-ph/0510256 

\bibitem[Laros et al.(1987)]{1987ApJ...320L.111L} Laros, J.~G., et al.\ 
1987, ApJL, 320, L111 

\bibitem[Lazzati et al.(2005)]{2005MNRAS.362L...8L} Lazzati, D., Ghirlanda, 
G., \& Ghisellini, G.\ 2005, MNRAS, 362, L8 

\bibitem[Levan et al.(2006)]{2006MNRAS.368L...1L} Levan, A.~J., Wynn, 
G.~A., Chapman, R., Davies, M.~B., King, A.~R., Priddey, R.~S., \& Tanvir, 
N.~R.\ 2006, MNRAS, 368, L1 

\bibitem[Lin et al.(2003)]{2003AJ....126.1286L} Lin, W., et al.\ 2003, AJ, 126,
1286

\bibitem[Lipunov et al.(2005)]{2003GCN..4206....1O} Lipunov, V., et al. 2005, GRB Coordinates Network, 4206, 1 

\bibitem[Lithwick \& Sari(2001)]{2001ApJ...555..540L} Lithwick, Y., \& 
Sari, R.\ 2001, ApJ, 555, 540 

\bibitem[Matonick \& Fesen(1997)]{1997ApJS..112...49M} Matonick, D.~M., \&
Fesen, R.~A.\ 1997, ApJS, 112, 49

\bibitem[Monet et al.(2003)]{2003AJ....125..984M} Monet, D.~G., et al.\ 
2003, AJ, 125, 984

\bibitem[Nakar \& Piran(2002)]{2002MNRAS.330..920N} Nakar, E., \& Piran, 
T.\ 2002, MNRAS, 330, 920 

\bibitem[Nakar et al.(2006)]{2006ApJ...640..849N} Nakar, E., Gal-Yam, A., 
Piran, T., \& Fox, D.~B.\ 2006, ApJ, 640, 849 
 
\bibitem[Nakar et al.(2005a)]{2005astro.ph.11254N} Nakar, E., Gal-Yam, A., 
\& Fox, D.~B.\ 2005a, astro-ph/0511254 

\bibitem[Nakar et al.(2005b)]{2005ApJ...635..516N} Nakar, E., Piran, T., \& 
Sari, R.\ 2005b, ApJ, 635, 516 

\bibitem[Ofek(2006)]{2006ApJ...N} Ofek, E.~O. 2006, in prep.

%\bibitem[Ofek et al.(2005)]{2005GCN..4208....1O} Ofek, E.~O., Cenko, S.~B., 
%Soderberg, A.~M., Kulkarni, S.~R., \& Fox, D.~B.\ 2005, GRB Coordinates 
%Network, 4208, 1 

\bibitem[Palmer et al.(2005)]{2005Natur.434.1107P} Palmer, D.~M., et al.\ 
2005, Nature, 434, 1107 

\bibitem[Parsons et al.(2005)]{2005GCN..4363....1P} Parsons, A., et al.\ 
2005, GRB Coordinates Network, 4363, 1 

\bibitem[Perelmuter \& Racine(1995)]{1995AJ....109.1055P} Perelmuter, 
J.-M., \& Racine, R.\ 1995, AJ, 109, 1055

\bibitem[Perez-Gonzalez et al.(2006)]{2006astro.ph..5605P} P\'{e}rez-Gonz\'{a}lez,
P.~G., et al.\ 2006, ArXiv Astrophysics e-prints, astro-ph/0605605

\bibitem[Petit et al.(1988)]{1988A&AS...74..475P} Petit, H., Sivan, J.-P.,
\& Karachentsev, I.~D.\ 1988, A\&AS, 74, 475

\bibitem[Popov \& Stern(2005)]{2005MNRAS.tmp.1092P} Popov, S.~B., \& Stern, 
B.~E.\ 2005, MNRAS, 1092 

\bibitem[Schlegel et al.(1998)]{1998ApJ...500..525S} Schlegel, D.~J., 
Finkbeiner, D.~P., \& Davis, M.\ 1998, ApJ, 500, 525 

%\bibitem[Schmidt et al.(1993)]{1993AN....314..371S} Schmidt, K.-H., Priebe, 
%A., \& Boller, T.\ 1993, Astronomische Nachrichten, 314, 371 

\bibitem[Tanvir et al.(2005)]{2005Natur.438..991T} Tanvir, N.~R., Chapman, 
R., Levan, A.~J., \& Priddey, R.~S.\ 2005, Nature, 438, 991 
 
%\bibitem[Taylor et al.(2005)]{2005ApJ...634L..93T} Taylor, G.~B., et al.\ 
%2005, ApJL, 634, L93 

\bibitem[Terasawa et al.(2005)]{2005Natur.434.1110T} Terasawa, T., et al.\ 
2005, Nature, 434, 1110 

%\bibitem[Tiengo et al.(2005)]{2005A&A...440L..63T} Tiengo, A., Esposito, 
%P., Mereghetti, S., Rea, N., Stella, L., Israel, G.~L., Turolla, R., \& 
%Zane, S.\ 2005, \aap, 440, L63 

%\bibitem[Voges et al.(1999)]{1999A&A...349..389V} Voges, W., et al.\ 1999, 
%A\&A, 349, 389 

%\bibitem[Voges et al.(2000)]{2000IAUC.7432....3V} Voges, W., et al.\ 2000, 
%IAU Circ., 7432, 3 

\bibitem[Woods \& Thompson(2004)]{2004astro.ph..6133W} Woods, P.~M., \& 
Thompson, C.\ 2006, in Compact Stellar X-ray Sources, Eds. W.H.G. Lewin \& M. van~der~Klis, Cambrideg University Press, astro-ph/0406133 
 

\end{thebibliography}
\end{document}